\documentclass[a4paper]{jpconf}
\usepackage{graphicx}

\usepackage{hyperref}
\hypersetup{
    colorlinks=true,
    linkcolor=blue,
    filecolor=magenta,      
    urlcolor=blue,
    citecolor=red
}

\begin{document}
\title{\large Innovative approaches to high school physics competitions: Harnessing the power of AI and open science}

\author{D Borovský$^{1}$, J Hanč$^{1}$ and M Hančová$^{2}$}

\address{$^1$Institute of Physics, Faculty of Science, Pavol Jozef Šafárik University in Košice, Slovakia}
\address{$^1$Institute of Mathematics, Faculty of Science, Pavol Jozef Šafárik University in Košice, Slovakia}

\ead{jozef.hanc@upjs.sk$^1$}

\begin{abstract}
High school physics competitions serve as a platform for talented students to showcase their skills, engage in challenging problems, and foster a passion for science. This paper explores innovative approaches to enhance these competitions by harnessing the power of open science and artificial intelligence (AI) tools. Particularly we delve into the capabilities of state-of-the-art AI chatbots, i.e. ChatGPT, Bard, Claude, related to problem solving in physics. Together with open science tools like SageMath and Jupyter AI, they have the potential to serve as intelligent, powerful co-pilots, tutors, and assistants in understanding and applying physics, as well as knowledge from connected STEM fields. Furthermore, these innovative approaches can revolutionize high school physics competitions, providing students and their tutors with powerful resources to excel in their scientific pursuits.
\end{abstract}

\section{Introduction}
High school physics competitions and preparation for them are an essential part of informal physics education at the secondary level. Special places in these high school competitions are held by the Physics Olympiad, the PhO (\url{https://www.ipho-new.org}), and the Young Physicist Tournament, the YPT (\url{https://www.iypt.org}), both of which have long-standing traditions with international significance, ranking them among the most prestigious. In these competitions, talented students can not only refine and deepen their theoretical and experimental knowledge and skills but also compare with their peers. 

\noindent In the broader context, participation in the PhO or/and YPT not only provides an opportunity for competition but also has a strong impact on shaping the students' future careers. The successful participation often guides talented students not only for university studies in natural science and technical fields but equips them with a high level of critical thinking and scientific literacy essential also in the study of medicine, economics, or law. In other words, such competitions appear also important from the perspective of current trends in STEM education and the needs of our society, where STEM disciplines are the primary drivers of the economy and where our jobs, health, education $\&$ environment, and even roles as citizens become strongly dependent on STEM, digital, and data literacy.

\noindent With the ubiquitous rise and implementation of artificial intelligence (AI)  in recent years, particularly exemplified this year (2023) by phenomenal AI chatbots like ChatGPT (\url{https://chat.openai.com}), we have decided to explore the potential of such technologies in preparing students for competitions like PhO and YPT. In this article, we present our ongoing research and latest findings dealing with this task.

\section{Our Research and Methodology}

Our research concerning generative AI in physics education (which is a part of the Ph.D. study of DB) is connected to the following research questions:

\begin{itemize}
    \item \textit{How can state-of-art AI Chatbots, like ChatGPT, be integrated into physics education to support physics teaching and learning?}
    \item \textit{What are the perceptions of students and teachers concerning the effectiveness of the generative AI tools in enhancing understanding and engagement in physics?}
    \item \textit{How does the use of generative AI impact students' performance in physics, as quantified by empirical data?}
\end{itemize}

\noindent As for research design and methodology, to find answers we have decided to apply an exploratory sequential mixed-methods design \cite{creswellDesigningConductingMixed2017} with a focus on qualitative perspectives supported by quantitative data (design QUAL $\Rightarrow$ quan). Currently, in the qualitative phase, we are undertaking a literature review to understand the theoretical framework and potential of generative AI like ChatGPT in physics education. This qualitative stage also encompasses a pilot case study, where we are exploring the capabilities and performance of generative AI in diverse physics tasks, as well as its integration into physics education. One of our prime examples of such integration is the field of informal preparation for competitions like the PhO and YPT. The forthcoming planned quantitative phase will provide data to measure the impact of AI within the existing conditions of physics education.

\section{Technology and Models behind the State-of-Art AI Chatbots}

To uncover the capabilities and truly grasp the performance of generative AI tools in the context of current published works related to education, we cannot solely rely on direct interaction with ChatGPT through an experimental approach, as illustrated in references \cite{kortemeyerCouldArtificialintelligenceAgent2023a, westAIFCICan2023, dahlkemperHowPhysicsStudents2023, daherArtificialIntelligenceGenerative2023}. It's also crucial to delve into a more detailed theoretical analysis and understanding of what ChatGPT actually is, and how the model behind ChatGPT works from technology and mathematics perspective.

\noindent At the heart of today’s phenomenal chatbots, like ChatGPT \cite{openaiGPT4TechnicalReport2023}, lies a unique and revolutionary artificial neural network architecture known as the transformer. Introduced by researchers from Google in 2017 \cite{vaswaniAttentionAllYou2017}, the transformer model has since been the backbone of many breakthroughs in AI. We do not explain in detail how transformers work (explanations at different levels can be found in \cite{linSurveyTransformers2022a, zhaoHowChatGPTReally2023, wolframWhatChatGPTDoing2023}). For our purposes, it is sufficient to look at a transformer in analogy with a human brain.

\noindent As a type of neural network, transformer as a neural network is similar to the neurons network in our human brain. The size of transformers is given by a number of parameters and if a number of parameters is more than $10^9$ we call a transformer for natural language processing a \textit{Large Language Model} (LLM) \cite{zhaoSurveyLargeLanguage2023}. Every transformer has to be taught on a dataset after creation and initialization, similarly as a newborn has to learn about the world. This process is called \textit{pre-training}, in analogy with the human brain, it is like early learning when a baby learns basic concepts, sounds, visuals, languages, etc.  

\noindent After pre-training, the model is \textit{fine-tuned} (together with implementing ethical norms) on a smaller, specific dataset for particular tasks like translation, question-answering, or summarization. ChatGPT was fine-tuned with the aim of improving my performance for user interactions. Fine-tuning corresponds to schooling or formal education in human life which can be diverse and in the final stage is tailored towards specific content and skills of the chosen profession (career). A pre-trained and fine-tuned transformer designed for natural language processing is called a \textit{Generative Pre-trained Transformer} or GPT (for more accurate details on what type of transformer GPT exactly is, see \cite{linSurveyTransformers2022a}). 

\begin{figure}[h]
    \includegraphics[width=1\textwidth]{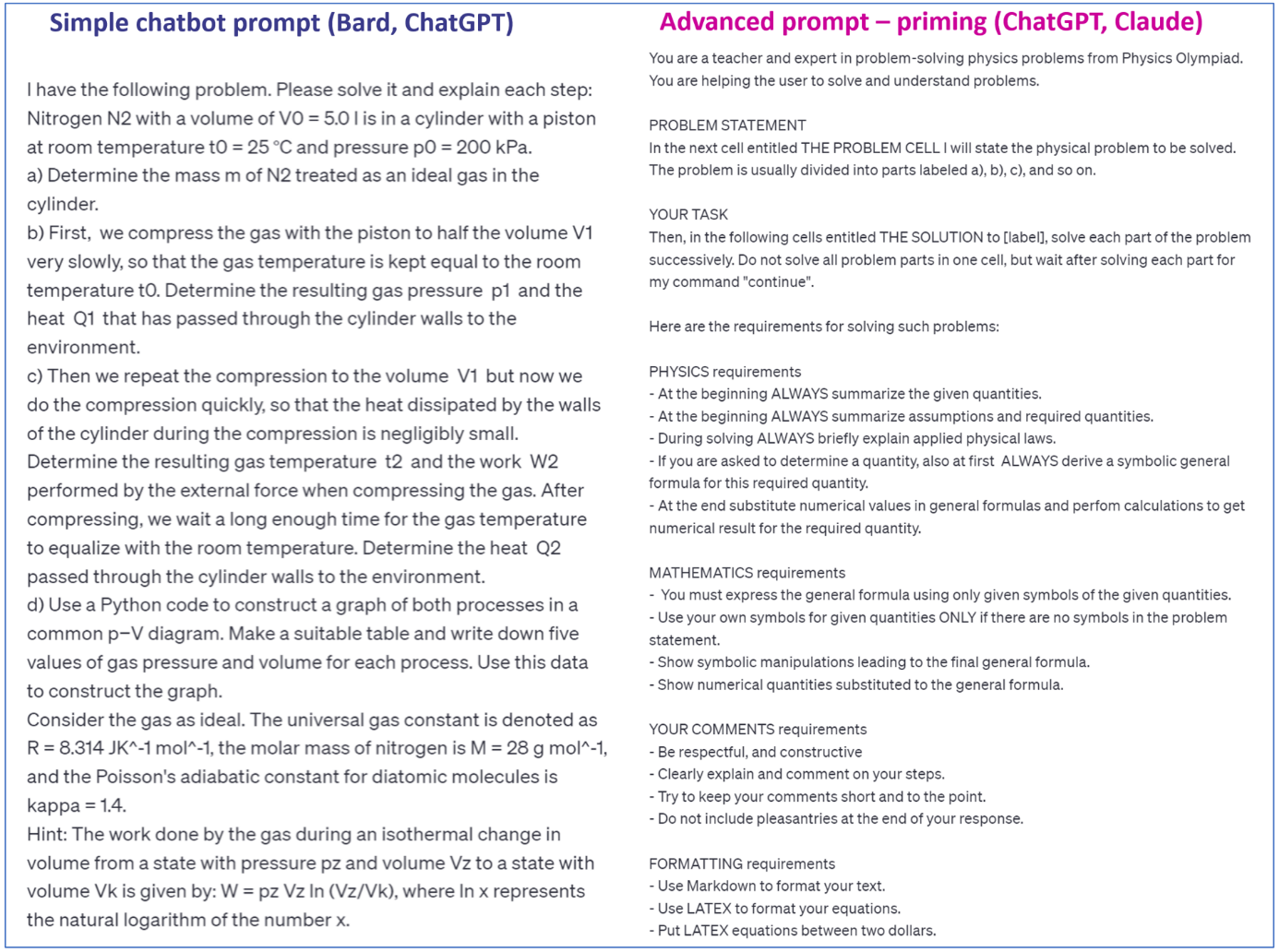}
    \caption{\label{fig1}The problem from the Physics Olympiad (2022, regional round, category C) as prompts in AI chatbots: a simple prompt (on the left) and with an improving advanced prompt using priming (on the right).}
\end{figure}

\noindent ChatGPT with model GPT-3.5 was trained on a gigantic and diverse dataset equivalent to reading 45 million books in 26 languages. The language capabilities of ChatGPT are therefore phenomenal compared to its predecessors. It can interactively communicate with you indistinguishably from a human, remembering the content of the conversation within a single session. It can discuss and draw conclusions on virtually any topic, including history, science, art, or technology. In interactions with a chatbot like ChatGPT, any text or question we pose, expecting a response, is called a \textit{prompt} (see Fig.~\ref{fig1}).

\noindent It's crucial to recognize that re-inputting the same prompt doesn't guarantee an identical response from the chatbot. This variability isn't a glitch, but it is intentional. In an LLM model, it is determined by a technical parameter called \textit{temperature}. By default, many chatbots set the temperature parameter at 0.7 for their large language models. When the temperature is set to zero, the chatbot's output remains the same, reflecting the highest probability response from the neural network structure. With any non-zero value, the chatbot has the possibility to randomly select a different answer with a probability close to the maximum. This means that the higher the temperature, the further from the optimal answer the chatbot can deviate. It also means that a higher temperature will result in less boring, more creative and unpredictable text. It is analogous to the concept of temperature in physics, where a higher temperature means the more chaotic and unpredictable motion of particles and vice versa.

\begin{figure}[h]
    \includegraphics[width=1\textwidth]{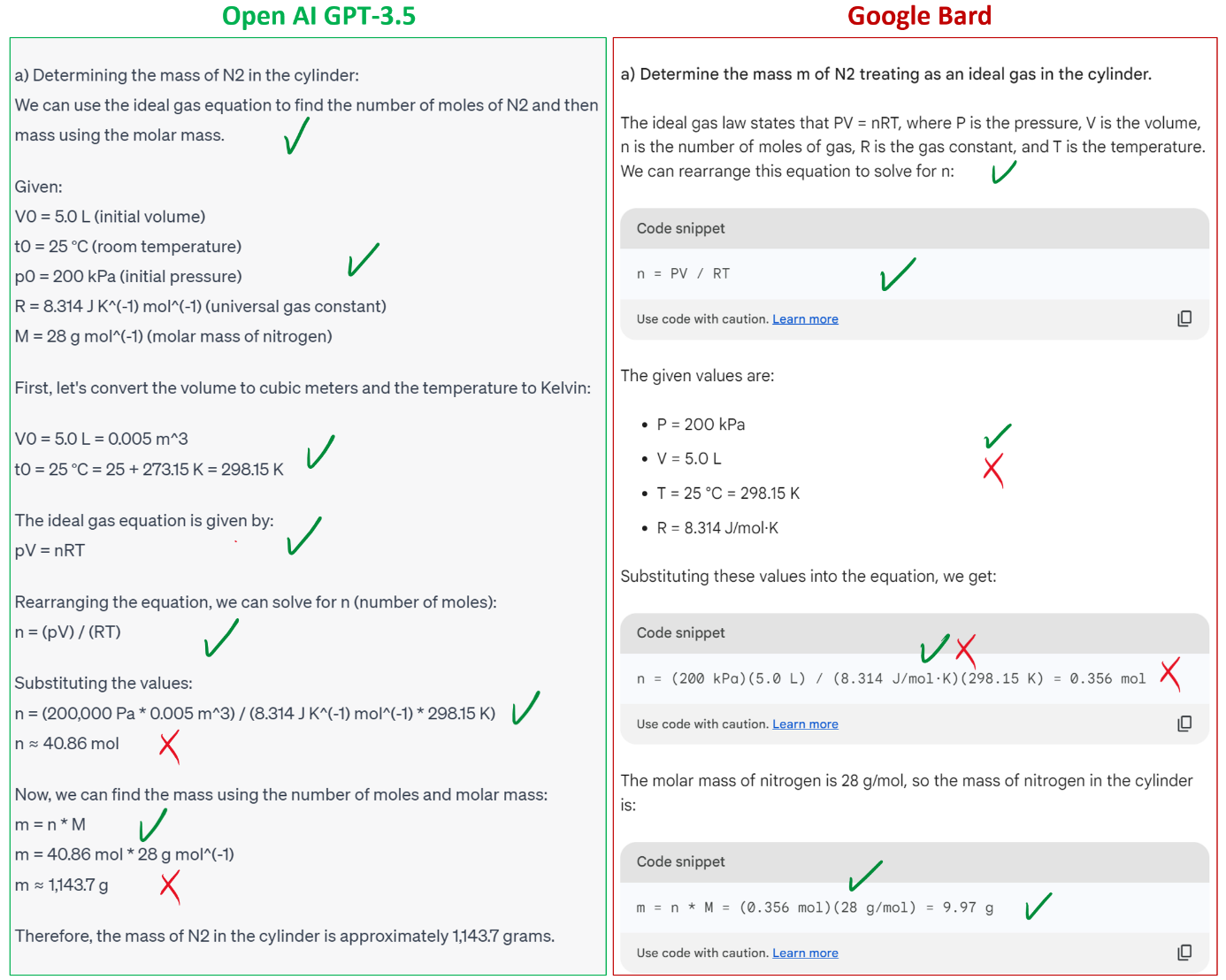}
    \caption{\label{fig2} Comparing solutions for part a) of the PhO problem from Fig.~\ref{fig1} provided by free versions of AI chatbots from OpenAI (GPT-3.5) and Google (Bard).}
\end{figure}

\noindent From a technological perspective, the release of ChatGPT in November 2022 wasn't so groundbreaking event, as numerous LLM transformers, including GPT-3, were already available (see Tab.1, Fig.~\ref{fig2} in \cite{zhaoSurveyLargeLanguage2023}). However, the release of ChatGPT became one of the milestones igniting accelerated research in the LLM field. Another important milestone was the release of LLM with a free, open license from Meta called LLaMA \cite{touvronLlamaOpenFoundation2023}.

\section{Current Capabilities of Generative AI Tools}

\subsection{Problem Solving with the State-of-Art AI Chatbots}

To determine the current capabilities of State-of-the-Art AI Chatbots in solving physics tasks and problems from PhO and YPT, we conducted a simple pilot case study. We have chosen currently the most powerful chatbots using the strongest LLMs (according to chatbot arena at \url{https://chat.lmsys.org}): 

\begin{itemize}
    \item the aforementioned \textit{ChatGPT} from OpenAI \cite{openaiGPT4TechnicalReport2023}, supported by Microsoft (with the free GPT-3.5 model and the paid GPT-4 model and plugins available at \url{https://chat.openai.com}) 
    \item \textit{Bard} from Google using PaLM2 \cite{anilPaLMTechnicalReport2023} (\url{https://bard.google.com}; via a Google account), 
    \item \textit{Claude} from the Google-backed Anthropic \cite{anthropicModelCardEvaluations2023}  (\url{https://claude.ai}; via the Opera with VPN),
    \item \textit{Vicuna}, the open-source chatbot \cite{zhengJudgingLLMasajudgeMTBench2023} from Berkeley University based on the LLaMA model from Meta (accessible in the chatbot arena \url{https://chat.lmsys.org}). 
\end{itemize}

\subsubsection{PhO Problem.}

As the first problem in the study, we chose a task from the 63rd edition of the Slovak Physics Olympiad (regional round, category C – 2nd year of high school) from the academic year 2021/2022 (authors: Bystrický \& Konrád). The task is shown in Figure 1 as a simple prompt. In Figure 2, we see the best solutions (chosen from 10 times repeatedly produced solutions to the same prompt) of part a) provided by OpenAI GPT-3.5 and Google Bard. Both chatbots were characterized by great speed (they generated the solution almost instantly). Additionally, Bard provided a preview of three different responses in the sense of the temperature parameter described earlier, from which we can choose and continue the conversation accordingly.

\noindent We see that both chatbots provided annotated structured solutions, with the correct application of the physical law. Such annotations are what we expect in solutions to PhO tasks. However, both chatbots made errors. Google Bard did not convert quantities into SI units, an oversight sometimes also seen among younger PhO solvers. In both cases, the chatbots failed in mathematics.

\noindent The correct values are $n = 0.403 ~\mathrm{mol}$, $m = 11.3 ~\mathrm{g}$ which means that GPT-3.5 value was 2 orders of magnitude higher. We'll just mention here that the Vicuna-33B chatbot solved the problem as competently as GPT-3.5 did but was significantly slower and also with numerical errors ($n = 0.219 ~\mathrm{mol}$, $m = 6.0698 ~\mathrm{g}$). None of the chatbots were able to correct numerical errors, even after multiple prompts.

\noindent Strictly speaking, from the perspective of the rules of the PhO, none of the chatbots demonstrated the required general solution. This behavior is understandable since we didn't prompt the chatbots to do so. In such cases, we can make further prompts iteratively to get the desired general solution. However, a challenge explaining general solutions is not a trivial matter. Based on our multiple interactions with chatbots and also the literature review, it seems that a very effective prompt strategy is called \textit{priming}. Generally, the right priming improves the performance of language models by providing them with context or initial information before generating text. This is typically done by supplying the model with a more advanced prompt or a set of input tokens related to the task or context in question. There are several priming techniques (see \cite{johnArtAskingChatGPT2023, zhaoSurveyLargeLanguage2023}), and an example is the advanced prompt in Fig.~\ref{fig1} (on the right). 

\begin{figure}[h]
    \includegraphics[width=1\textwidth]{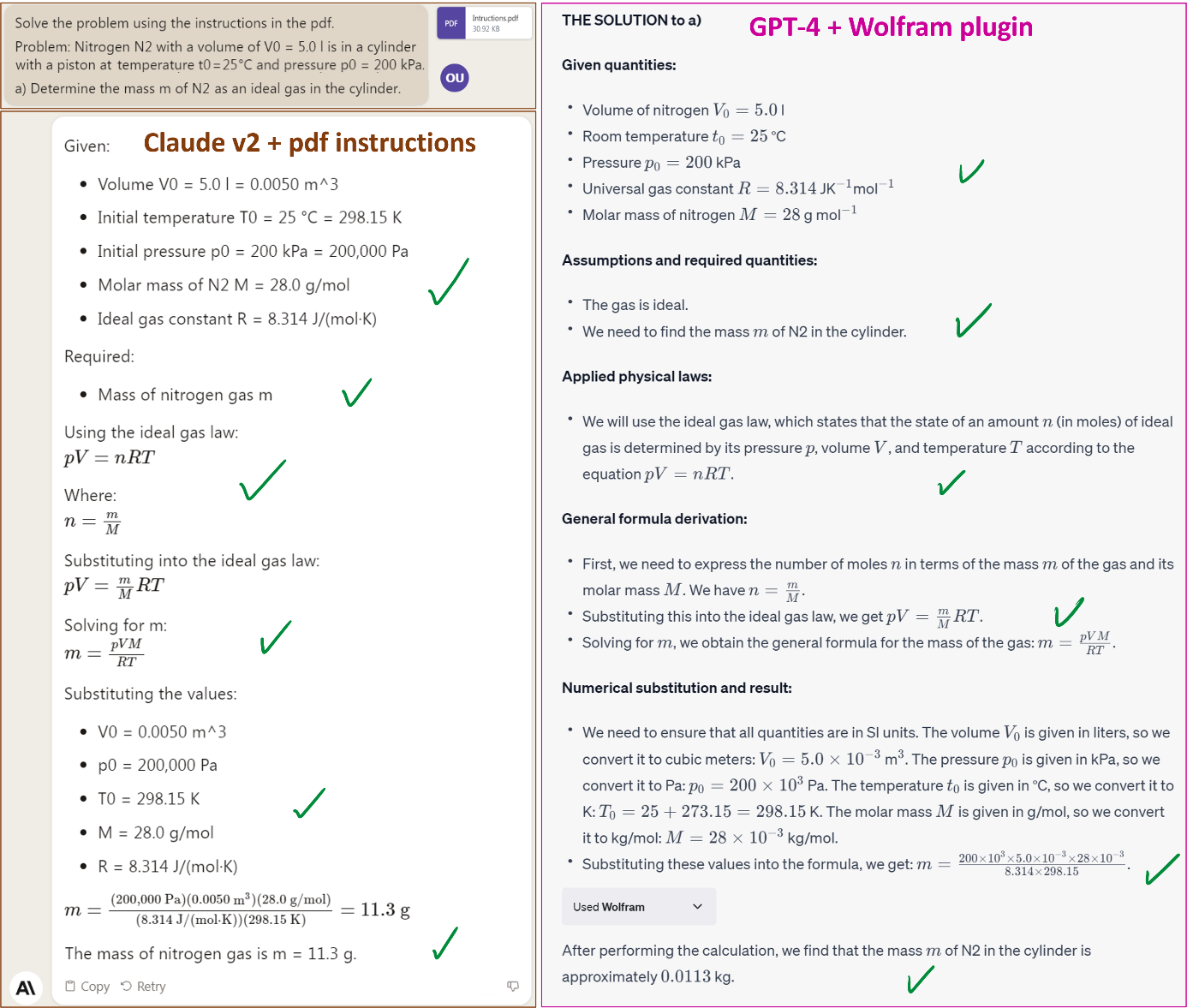}
    \caption{\label{fig3} The state-of-art AI chatbots solutions using priming and plugins: Claude v2 (on the left; with instructions as a pdf attachment) and GPT-4 solution (on the right; with plugin Wolfram).}
\end{figure}

\noindent In Fig.~\ref{fig3}, we can see the completely correct annotated general solution from Claude and ChatGPT, along with the correct numerical results. We want to highlight that, in the case of the Claude chatbot, it's possible to upload instructions in pdf form which simplifies and clarifies communication. Initially, Claude provided a numerically incorrect result ($m = 0.70 ~\mathrm{g}$), which he corrected only after our challenge.

\noindent In the case of ChatGPT, we see the solution in the paid GPT-4 model with the Wolfram plugin. This means that every numerical result is secretly calculated not by GPT-4, but by Wolfram Alpha, a computational knowledge engine developed by Wolfram Research (\url{https://www.wolframalpha.com}). Thanks to this plugin, ChatGPT almost certainly provides correct numerical results. In rare cases, the Wolfram plugin fails, gets stuck in a loop, and provides no answer. In such situations, it is necessary to solve the problem from scratch.

\subsubsection{YPT Problem.}

From the perspective of physical complexity, task a) was of a basic knowledge level related to the ideal gas theme, which even an average high school student should manage. In our second instance, we tackled a much more challenging problem from the YPT competition (problem 12 from the 34th IYPT). The problem's statement was as follows:

\begin{quote}
\textit{A Wilberforce pendulum consists of a mass hanging from a vertically oriented helical spring. The mass can both move up and down on the spring and rotate about its vertical axis. Investigate the behavior of such a pendulum and how it depends on relevant parameters.}
\end{quote}

\noindent We were able to solve the problem, concerning its dynamics, successfully with the Chatbot ChatGPT using its two paid models (applying an advanced modelling approach -- Lagrange analytical method; here it is recommended to install a browser extension \cite{michaelTeXAllThings2023} for rendering LaTeX equations): 
\begin{itemize}
    \item GPT-4 model: \\ \url{https://sharegpt.com/c/MZPHMS2}
    \item GPT-4 code interpreter: \\ \url{https://chat.openai.com/share/0d059dc6-1632-4e9f-af46-4c5926b91304}
\end{itemize}

\subsection{Generative AI in Open Science Jupyter Notebooks}

Problems faced by AI in using mathematics, whether related to the accuracy of numerical results or algebraic manipulations, can be addressed by integrating AI with open science tools designed for scientific computing and data processing. Moreover, this integration also facilitates a more straightforward and efficient interaction with AI during scientific computing.

\noindent One of such open science tools is a free Python-based mathematics software called SageMath (\url{https://www.sagemath.org}). In our previous works \cite{gajdosInteractiveJupyterNotebooks2022b, borovskyScientificComputingOpen2023a} we introduced and described the capabilities of SageMath in research and STEM education. We also demonstrated that SageMath is ideally suited for solving PhO and YPT problems. Technologically, from a visual standpoint, SageMath is a \textit{Jupyter Notebook} -- an interactive web page that runs on any modern web browser and can be easily edited as a Word document. It allows users to insert text, images, and videos, as well as computational code representing variables, equations, or commands for plotting graphs. Moreover, any Python library or extension designed for Jupyter can be installed and utilized in SageMath. In this article, we show two possibilities for integrating AI into Jupyter that we identified through our research in relevant resources.

\subsubsection{AI Assistant Jupyternaut.}
In March 2023, also a team from Project Jupyter (\url{https://jupyter.org}) began intensively working on AI implementation through a sub-project named Jupyter AI \cite{projectjupyterJupyterAIJupyter2023}. The culmination of this effort is the \verb"jupyter_ai" extension, which integrates the Jupyter notebook with AI, allowing users to choose from various large language models (several dozens), including the OpenAI GPT-3.5, which was chosen by us. Upon its installation, an AI chatbot named Jupyternaut appears. 

\begin{figure}[h]
    \includegraphics[width=1\textwidth]{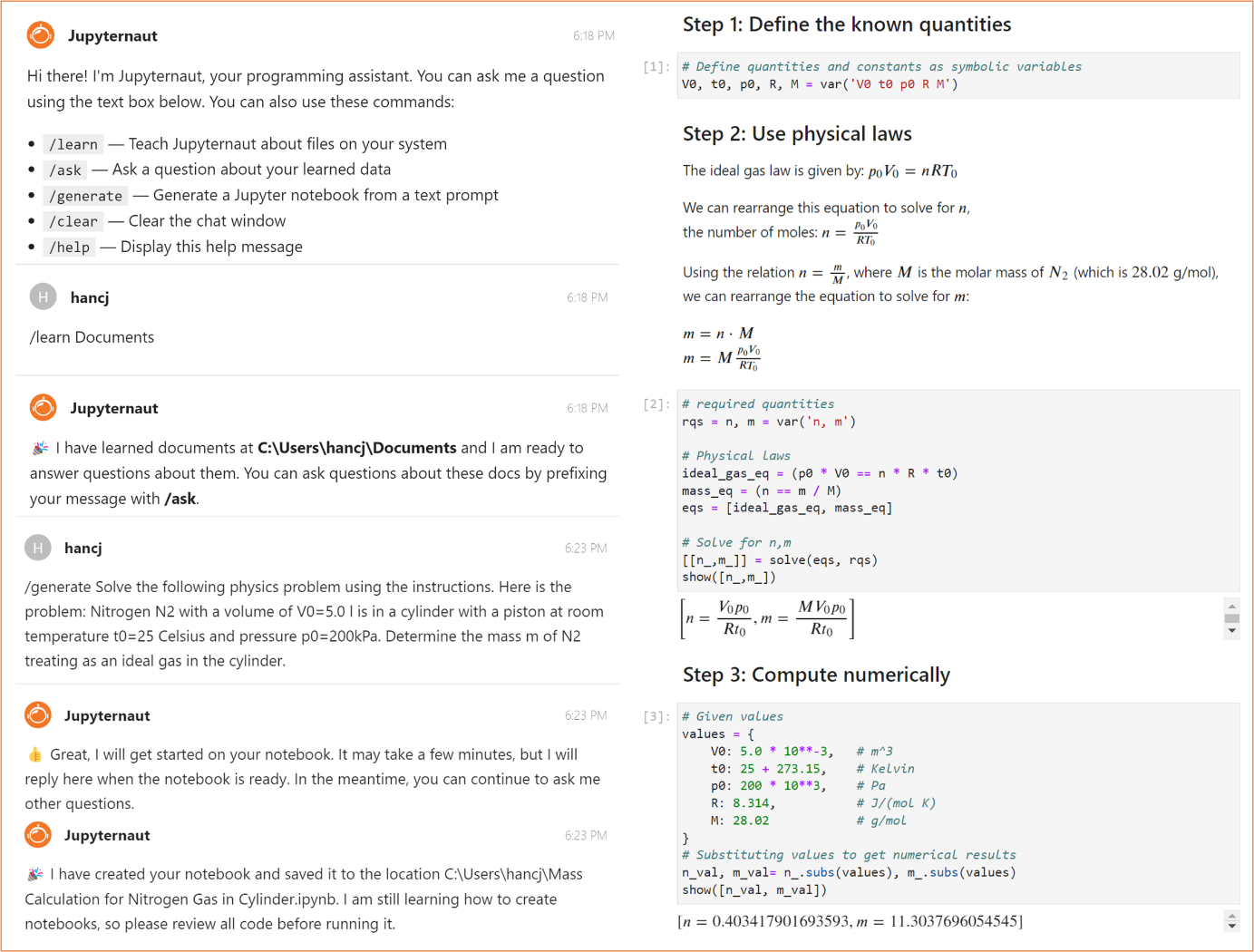}
    \caption{\label{fig4} A solution generated by Jupyternaut (with GPT-3.5) with corresponding conversation (on the left) and the resulting SageMath Jupyter notebook (on the right).}
\end{figure}

\noindent This chatbot acts as a general-purpose AI virtual assistant -- serving mainly for coding purposes and, as for physics (or natural sciences), becomes a helpful guide for computations, explaining and applying physics concepts, and laws. Remarkably, with the aid of GPT-3.5, it can do so in 26 languages, including Slovak. Figure 4 (on the left) illustrates the AI's interaction within SageMath, showing a conversation during solving our PhO problem. Using the \verb"/learn" command, we requested Jupyternaut to study all materials within the \textit{Documents} folder. In our case, it was a text with general instructions for solving physics problems (Fig.~\ref{fig1}, on the right), complemented by guidelines stating that explanations should be rendered in SageMath as text cells, while calculations and symbolic manipulations should be done as computation cells. Then, with the \verb"/generate" command, we asked Jupyternaut to create a Jupyter notebook as SageMath, tasked with solving the given physics problem. In a brief moment, the virtual assistant produced a notebook, which after our minor adjustments is displayed in Fig.~\ref{fig4} (on the right).

\noindent Now, we can see the general solution, with numerical results, accurate and precise up to 15 decimal places. By highlighting any section of the notebook on the right, Jupyternaut can provide a detailed explanation (tailored to our prompts) or fix errors. 

\noindent Employing an AI assistant directly within a digital scientific environment as SageMath, compared to a web-based chatbot interface, is substantially more straightforward. The need for copying content back and forth is eliminated. Concurrently, the synergy with SageMath gives us better interactivity at intermediate steps, easy modifiability, reproducibility, and direct visualization capabilities.

\subsubsection{ChatGPT – Jupyter – AI Assistant.}
A second, very easy solution for bringing AI into the SageMath Jupyter notebook is through the use of a browser extension, named \textit{ChatGPT -- Jupyter -- AI Assistant} \cite{kokChatGPTJupyter2023a}, which is compatible with browsers like Chrome, Edge, or Opera. This extension using only GPT-3.5 can be installed with a single click, appearing as a toolbar with various control buttons within the Jupyterlab environment. Similar to Jupyternaut, it can comment on the highlighted computational code, format it, explain it, fix its errors, and also generate new code based on the prompt.

\section{Conclusion}

In this paper, we explored the role and capabilities of generative artificial intelligence through state-of-art chatbots like ChatGPT to enhance high school physics competitions. The results indicate the undeniable benefits of using chatbots in the context of the Physics Olympiad (PhO) and Young Physicist Tournament (YPT). These AI tools, when combined with open science platforms like SageMath, can remove problems with the mathematics that even the best chatbots struggle with alone. They could also guide students through intricate calculations, offer explanations for physical phenomena, and suggest various approaches to tackle a given problem, while ensuring every step is transparent and reproducible.

\noindent However, over-reliance on chatbots may inadvertently reduce a student's critical thinking and problem-solving skills. Since general-purpose language models like ChatGPT are also pre-trained on a large body of Internet data they produce misconceptions and AI hallucinations. To address these concerns, such AI tools must be adapted in the context of active learning to be facilitators and not authorities. An example of such adaptation is experimentally tested Khanmigo, a fine-tuned version of ChatGPT \cite{khanSalKhanHow}. While Khanmigo has been created to suit educational needs in math teaching, we believe that thanks to rapid progress fine-tuned open-source language models will serve specifically to the demands of physics education and competitions.

\noindent In a world where STEM and digital literacy are becoming increasingly crucial, merging cutting-edge AI tools with traditional methods of education might just be the perfect recipe for nurturing the next more powerful generation of scientists and researchers. 

\section{References}
\bibliographystyle{iopart-num}
\bibliography{DIDFYZ2023}

\section*{Acknowledgments}
This work is supported by the Slovak Research and Development Agency under the Contract no. APVV-22-0515, APVV-21-0216 and APVV-21-0369.

\end{document}